\documentclass[aps,twocolumn,amsmath,amssymb,%
                prl,showpacs,showkeys]{revtex4}
\usepackage[dvips]{graphicx}
\usepackage{psfrag}

\begin{document}

\title{Spin--rotation interaction in fullerite C$_{60}$}

\author{V.P.~Tarasov}
\thanks{Corresponding author}
\email{tarasov@igic.ras.ru}
\author{Yu.B.~Muravlev}
\affiliation{Kurnakov Institute of General and Inorganic
Chemistry, Russian Academy of Sciences, Leninskii pr.~31, Moscow,
119991 Russia}

\author{D.E.~Izotov}
\email{izotov@mpipks-dresden.mpg.de}%
\affiliation{Max Planck Institute for the Physics of Complex
Systems, Noethnitzer str. 38, Dresden, 01187 Germany}

\begin{abstract}
We report on the $^{13}$C spin--lattice relaxation times in
polycristalline fullerite C$_{60}$ measured over the temperature
range 295~K to 1000~K. At temperatures above 470~K, spin--lattice
relaxation is dominated by the spin--rotation interaction. From
the analysis of temperature dependence of $T_1(^{13}$C), the
spin--rotation constant is determined using {\it J-\/} and {\it
M-}models of rotational diffusion: $C_J\simeq -51$~Hz, $C_M\simeq
-29$~Hz.
\end{abstract}

\maketitle

\section{Introduction}
When a molecule rotates, the motion of its nuclear and electronic
charges produces alternative magnetic fields. These fields may
then interact with nuclear spins in the molecule. The resulting
spin--rotation interaction is one of the possible competing
contributions to the mechanisms of nuclear spin--lattice
relaxation in the gas and condensed phases~\cite{Ramsey53}. The
extraction of the contribution due to the spin-rotation mechanism
solely from the spin-lattice relaxation is of interest in several
ways. Firstly, it is especially useful to have a knowledge of the
molecular reorientation correlation time $\tau_\theta$ and the
angular momentum correlation time $\tau_J$ in order to
differentiate between alternative models of molecular rotation,
i.e., the rotational jump model versus the rotational diffusion
one. While $\tau_\theta$ can be determined reliably by a variety
of methods, $\tau_J$ is far less available to precise measurement.
The fact that the interaction between nuclear spin and magnetic
field generated by molecular rotation is a dominant mechanism for
spin--lattice relaxation provides an excellent approach to the
determination of $\tau_\theta$. Secondly, the spin--rotation
constants can be used to find the paramagnetic part of the
shielding tensor, and vice versa.

Fullerite C$_{60}$ is the only known solid that consists of
quasispherical homonuclear molecules. At ambient temperatures,
these molecules rotate quite fast, despite a very large moment of
inertia, $10^{-36}$~g~cm$^{2}$. Although the spin--rotation
constant in C$_{60}$ is relatively small (25.6~Hz, as estimated
from chemical shift), the spin--rotation interaction in this
compound would be expected to significantly contribute to
spin--lattice relaxation even at temperatures about
400~K~\cite{IzoPhCh02}.

$^{13}$C NMR line shape and spin--lattice relaxation in fullerite
C$_{60}$ have been studied in magnetic fields of 1.4 to 9.4~T at
temperatures ranging from 100 to
340~K~\cite{Yannon91,Tycko91,Mizo93,Prival97eng}. These
experiments clearly demonstrate that the $^{13}$C~NMR spectrum of
fullerite C$_{60}$ in the range 100--340~K is dominated by one
type of magnetic interaction, namely, by $^{13}$C magnetic
shielding anisotropy, which de\-ter\-mi\-nes both the line shape
and $T_1$. We have recently described rotational dynamics in
fullerite C$_{60}$ in terms of multiaxial discrete reorientations
of C$_{60}$ molecules and suggested that the experimental
manifestation of spin--rotation interaction C$_{60}$ should be
noticeable at temperatures above 400~K ~\cite{IzoPhCh02}.

Herein, we present the data in favor of this hypothesis. We report
on the results of measuring spin--lattice relaxation time
$T_1(^{13}$C) and $^{13}$C~NMR chemical shift in polycrystalline
fullerite C$_{60}$ at temperatures ranging from 295 to 1000~K and
on the experimental manifestation of spin--rotation interaction.
From theoretical analysis of the temperature behavior of
$T_1(^{13}$C), we determined correlation times of the C$_{60}$
rotation angular momentum and the spin--rotation coupling constant
of $^{13}$C nuclei in a fullerite C$_{60}$ molecule.

\section{Experimental}
A polycrystalline C$_{60}$ sample was obtained by crystallization
from a toluene solution, washed with hexane, and annealed in
vacuum at 200$^\circ$C. HPLC showed that the content of oxides and
higher fullerenes was less than 0.02\%. XRD confirmed a high
degree of crystallinity of the sample. The heat capacity of the
same sample was measured by DSC in the range
285--675~K~\cite{Dikii00eng}. The values obtained were consistent
with available data.

\begin{figure}[htb]
\centering
\psfrag{M}{{\small $M$, a.u.}}
\psfrag{t}{{\small $\tau_2$, s}}
\includegraphics[width=0.35\textwidth]{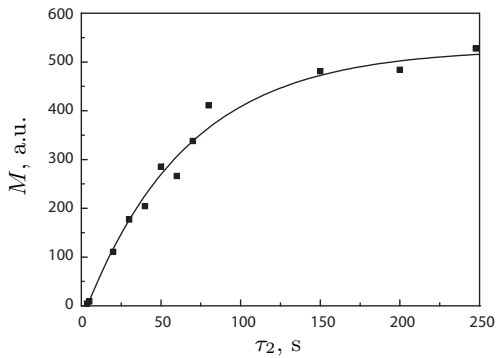}\\
\caption{Recovery of magnetization $M$ (arb. units) of $^{13}$C
spins in fullerite C$_{60}$ at 863~K after the action of a
saturating pulse train. The solid curve is described by the
equation $M(\tau)=M_\infty[1-\exp(-\tau/T_1)]$ at $T_1=59\pm5$~s.}%
\label{fig:rel}
\end{figure}

A C$_{60}$ powder (400~mg) was loaded into a quartz NMR tube 5 mm
in o.d. and 50 mm in length, evacuated to a residual pressure of
0.1~mm Hg, and sealed. NMR measurements were performed on a Bruker
MSL-300 spectrometer in a field of 7.04~T (75.4~MHz); an original
high-temperature probe was used to cover the temperature range
295--1000~K. A furnace 50~mm long with the maximal outer diameter
of 20~mm was designed with the use of two coaxial quartz tubes. A
noninductively wound heating element was placed in between these
tubes. An RF coil was fixed at a zircon tube 50~mm in length, 7~mm
in o.d., and 5.8 mm~in i.d. The tube with a sample was positioned
in the zircon tube. For thermal insulation of the furnace, quartz
wool was used. At a sample temperature of 1000~K, the temperature
of the probe cover was no more than 50$^\circ$C. Forced cooling of
the RF circuit was not used. Temperatures were calibrated with the
use of a Chromel--Alumel thermocouple. The temperature
reproducibility with allowance for the heat gradient was
$\pm20^\circ$~C. NMR spectra were excited by a one-pulse sequence
at a dead time of 4~ms and a pulse repetition time of 80~s; the
number of scans, 100; scan width, 10~kHz; SI 16K and TD 16K.
Chemical shifts were measured from CHCl$_3$ (77~ppm with respect
to TMS) as the external reference. Spin--lattice relaxation times
were measured with the use of the saturation--recovery pulse
sequence, $(90_x-t_1-90_x)_n-\tau_2-90_y-5T_1$. The $\pi/2$ pulse
width was 9~ms, $n=100$, and time delays $\tau_2$ were from 1 to
500~s. Time $t_1=100$~ms was optimized so that a signal upon
saturation was absent. To construct the plots of signal amplitude
versus delay time, 14 to 17 $\tau_2$ values were used. The $T_1$
values were obtained through processing the amplitude--$\tau_2$
plots with the SIMFIT program (ASPECT--3000 software) with
allowance for one-exponential magnetization recovery.
Figure~\ref{fig:rel} shows a representative plot of signal
amplitude recovery.
Before each run, the sample was kept at a specified temperature
for no less than 60--100~min. It took about 5~h to measure $T_1$
at one temperature point. Before each run, the sample was allowed
to cool to room temperature. The error in measurement of $T_1$ was
7--10\%.

\section{Results}
\begin{figure}[htb]
\centering
\psfrag{t}{{\small $T$, K}}
\psfrag{s}{{\small $\delta$, ppm}}
\includegraphics[width=0.35\textwidth]{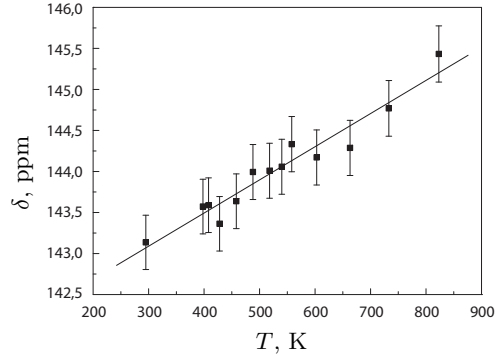}\\
\caption{$^{13}$C NMR chemical shift (ppm from TMS) in fullerite
vs. temperature. Regression line: $\delta=141.8+4\times10^{-3}T$.}
\label{fig:cs}
\end{figure}

\begin{figure}[htb]
\centering
\psfrag{T}{{\small $T$, K}}
\psfrag{t}{{\small $T_1$, s}}
\includegraphics[width=0.35\textwidth]{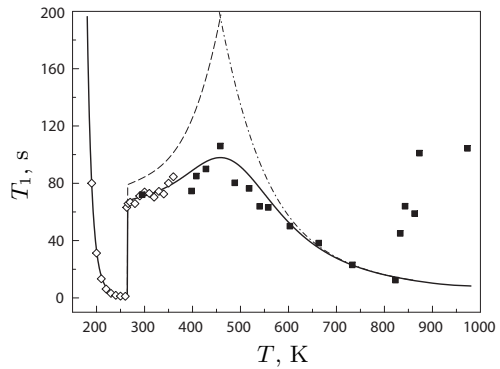}\\
\caption{Theoretical and experimental temperature de\-pen\-dence
of spin--lattice relaxation time $T_1(^{13}$C) in fullerite
C$_{60}$ at 190--1000~K ($B_0=7.04$~T). Solid line: Calculated
$T_1$. Dashed line: Contribution chemical shift anisotropy to
$T_1$. Dot--dash line: Spin--rotation contribution. Symbols
$\diamond$~: experimental data of Tycko et al.~\cite{Tycko91} and
Privalov et al.~\cite{Prival97eng}; {\tiny$\blacksquare$}~:
present work. Jump of $T_1$ at 260~K
corresponds to the reversible s.c.---f.c.c. phase transition.}%
\label{fig:T1all}
\end{figure}

The $^{13}$C NMR signal of the sample studied at 295~K had a
slightly asymmetric contour with a chemical shift of
$143.6\pm0.2$~ppm and the line width at half-maximum of 3.6~ppm,
which is consistent with the known
data~\cite{Yannon91,Tycko91,Mizo93,Prival97eng}. With an increase
in temperature to 800~K, the chemical shift linearly changes by
only 2.5~ppm (Fig.~\ref{fig:cs}) and the line width remains
unaltered within the error of measurements and is presumably
dominated by the field $B_0$ inhomogeneity.
In contrast to the line width and chemical shift, the
spin--lattice relaxation time $T_1(^{13}$C) experiences noticeable
changes in the temperature range studied (Fig.~\ref{fig:T1all}).
Increasing temperature from 295 to 473~K is accompanied by an
expected slight increase in $T_1$ from 77 to
107~s~\cite{IzoPhCh02}. A further increase in temperature results
in a smooth decrease in $T_1$ to 12~s at 823~K, being a strong
evidence that the dominating mechanism of $^{13}$C spin--lattice
relaxation in fullerite changes with an increase in temperature
from the magnetic shielding anisotropy mechanism at ambient and
low temperatures to the spin--rotation mechanism at high
temperatures. To the best of our knowledge, this is the first
example of realization of spin--rotation interaction in
solid-state $^{13}$C NMR. At temperatures 830--900~K, $T_1$
sharply increases to 100~s. In the range 900--1000~K, $T_1$ is
roughly temperature independent. The reasons behind this behavior
of $T_1(^{13}$C) at temperatures ranging from 830 to 1000~K are
not clear. Note that the changes in $T_1$ observed are reversible.

\section{Discussion}
\subsection{Chemical shift}
The linear change in chemical shift observed with a change in
temperature (Fig.~\ref{fig:cs}) is caused by averaging $^{13}$C
magnetic shielding over the rovibrational states of C$_{60}$
molecule
\begin{equation}
\langle\sigma\rangle\approx \sigma_\mathrm{e}+
2\sigma_s'\langle\Delta r_s\rangle+ \sigma_d'\langle\Delta
r_d\rangle,
\label{eq:sigma}
\end{equation}
where $\sigma_\mathrm{e}$ is the chemical shift constant at
equilibrium state, $\Delta r=r-r_\mathrm{e}$ is the change in
double ({\it d\/}) and single ({\it s\/}) bond lengths in
C$_{60}$, $\sigma_s'=(\partial\sigma/\partial r_s)_\mathrm{e}$,
$\sigma_d'=(\partial\sigma/\partial r_d)_\mathrm{e}$, and
$\langle\cdot\rangle$ denotes averaging over rovibrational states.
In combination with the data on bond lengths in C$_{60}$ at
temperatures 4--295~K~\cite{Lecler93}, Eq.~\eqref{eq:sigma} leads
to the following relation: $-400=-12.4\sigma_s'+3.5\sigma_d'$. If
we suppose that $\sigma_d'$ in C$_{60}$ is close to the one in
ethylene (Table~\ref{tab:sigma}), then
$\sigma_s'\simeq-20$~ppm~\AA$^{-1}$, which is comparable with the
magnitude of $\sigma_s'$ in ethane.
\begin{table}[htb]
\caption{C--C bond lengths, chemical shifts of $^{13}$C nuclei
with respect to bare $^{13}$C, and their derivatives in ethane,
ethylene, acetylene, and fullerene.} \label{tab:sigma}
\begin{ruledtabular}
\begin{tabular}{llll}
& $r$(300~K) & $\sigma$(300~K) & $(\partial\sigma/\partial
r)_\mathrm{e}$ \\
& (\AA) &  (ppm) & (ppm~\AA$^{-1}$) \\
\hline
C$_2$H$_6$ & 1.545 &  183.1\footnotemark[1]  & $-15$\footnotemark[1]  \\
C$_2$H$_4$ & 1.335 &  66.76\footnotemark[1]  & $-188$\footnotemark[1] \\
C$_2$H$_2$ & 1.205 &  121.35\footnotemark[1] & $-110$\footnotemark[1] \\
C$_{60}$   & 1.45, 1.40\footnotemark[2] &  61.0  & $-20$, $-188$\footnotemark[2]\footnotemark[3] \\
\end{tabular}
\end{ruledtabular}
\footnotetext[1]{C.~Jameson and H.~Osten, \emph{Theoretical
aspects of isotope effects on nuclear shielding. Annual reports on
{NMR} spectroscopy}, vol.~17 (Academic Press, London, 1986).}%
\footnotetext[2]{The numbers refer to single and double bonds,
respectively.}%
\footnotetext[3]{Estimated according to the relationship
$-400=-12.4\sigma_p'+3.5\sigma_h'$ (see comments in the text).}
\end{table}

\subsection{Spin--lattice relaxation time}
The experimental temperature dependence of $T_1$ can be adequately
described in the model of molecular
reorientations~\cite{IzoPhCh02}. This model reproduces $T_1(T)$
for both the low- and high-temperature phases up to 370~K, for
which the experimental data are known. In the low-temperature
simple cubic phase, C$_{60}$ molecules are assumed to jump between
symmetry-equivalent positions about molecular symmetry axes, while
in the high-temperature f.c.c. phase, reorientations take place
also about crystal symmetry axes.

For the jump model, $T_1$ due to chemical shift ani\-so\-tro\-py
is expressed in terms of correlation times for each type of
reorientation axes~\cite{IzoPhCh02}. With reasonable accuracy,
this formula can be approximated by the formula for the diffusion
rotation model with the effective correlation time $\tau_\theta$:
\begin{equation*}
(1/T_1)_{CS}=\frac{3}{10}\omega_0^2\delta^2
\left(1+\frac{\eta^2}{3}\right)
\frac{\tau_\theta}{1+\omega_0^2\tau_\theta^2},
\end{equation*}
where $\delta=\sigma_{33}-\sigma$, $\sigma=\sum_i\sigma_{ii}/3$,
$\eta=(\sigma_{11}-\sigma_{22})/\delta$ is the asymmetry
parameter, $\sigma_{ii}$ are the principal components of the
shielding tensor, and $\omega_0$ is the Larmor frequency. Values
of these parameters for fullerite C$_{60}$ were determined
earlier~\cite{Yannon91,Tycko91,Mizo93,Prival97eng}:
$\sigma=1.43\times10^{-4}$ (from TMS), $\delta=-1.1\times10^{-4}$,
$\eta=0.24$. Here, we use $\omega_0=2\pi\times 75.4$~MHz.

The decrease in $T_1$ observed at 470--820~K can be assigned to
the spin--rotation interaction, which becomes dominant in this
temperature range. This relaxation mechanism is characterized by
the angular momentum correlation time
$\tau_J$~\cite{Hubbard63,Gordon66,McClung69}. In terms of
so-called {\it J}- and {\it M}-models for rotational diffusion and
in the limit $\tau_J\ll\tau_\theta$, $\tau_J$ and $\tau_\theta$
for a spherical top molecule are related to each other as
$\tau_\theta\tau_J=I/6kT$ and $\tau_\theta\tau_J=I/2kT$ for {\it
J}- and {\it M}-models, respectively (where $I$ is the moment of
inertia of the molecule). The spin--lattice relaxation rate due to
spin--rotation interaction is given by
\begin{equation*}
(1/T_1)_{SR}=\frac{8\pi^2kT}{\hbar^2}IC^2\tau_J,
\end{equation*}
where $C$ is the spin--rotation constant.

The resulting relaxation rate is
$1/T_1=(1/T_1)_{CS}+(1/T_1)_{SR}$. At temperatures lower than
350~K, the relaxation rate is dominated by the chemical shift
anisotropy. Fitting the model data to the experimental data in
this temperature range gives the kinetic parameters of molecular
reorientations (Table~\ref{tab:param}). These parameters are
supposed to be also valid at higher temperatures. It is easy to
verify that, in this case, $\tau_J<\tau_\theta$ at temperatures
below 900~K and, hence, the relation between these two quantities
holds.
\begin{table}[htb]
\centering%
\caption{Kinetic parameters of molecular reorientations in
fullerite C$_{60}$ at 190--830~K in Arrhenius approximation. In
the low-temperature simple cubic phase the molecules are assumed
to jump about molecular symmetry axes. In the high-temperature
f.c.c. phase reorientations about crystal symmetry axes are added.
\label{tab:param}}
\begin{ruledtabular}
\begin{tabular}{llll}
$\tau_{0,\mathrm{mol}}$ & $E_\mathrm{mol}$ & $\tau_{0,\mathrm{cr}}$
& $E_\mathrm{cr}$ \\
($\times 10^{-14}$~s) & (kJ~mol$^{-1}$) & ($\times
10^{-10}$~s) & (kJ~mol$^{-1}$) \\
\hline 0.85 & 27.45 & 0.10 & 0.83
\rule[0pt]{0pt}{1.2em}\\
\end{tabular}
\end{ruledtabular}
\end{table}
The constant $C$ can be readily determined from the position of
the maximum of $T_1$ at 360~K, where $(T_1)_{CS}=(T_1)_{SR}$.
Since $I(\mathrm{C}_{60}) = 10^{-36}$~g~cm$^{2}$, we have
$|C_J|\simeq51$~Hz, $|C_M|\simeq 29$~Hz for $J$- and $M$- models,
respectively. The value of the spin--rotation constant obtained in
the $M$ model is in reasonable agreement with that derived from
the chemical shift, $-25.6$~Hz.

The model of molecular dynamics in fullerite C$_{60}$ suggested
adequately describes $T_1(T)$ in the range 190--830~K. However,
the most intriguing is the reversible effect associated with a
sharp increase in $T_1$ at 830--900~K, followed by the constancy
of $T_1$ in the range 900--1000~K. One of the possible reasons can
be a high-temperature phase transition~\cite{Peimo94}, although
high-temperature {\it in situ\/} studies of a C$_{60}$ powder by
high-resolution XRD did not disclose a deviation from linearity
for the volume expansion coefficient~\cite{Vogel96,Fis93}. This
problem calls for further investigation. We hope to address this
problem in the near future.

\begin{acknowledgments}
The high-temperature probe was manufactured by A.A. Samoilenko
(Radioelectronic optical laboratory), which is gratefully
acknowledged.
\end{acknowledgments}

\bibliographystyle{apsrev}

\end{document}